%% file: aph.tex
\documentclass[preprint2]{aastex}

\newcommand{\kms}{km s$^{-1}\;$}
\newcommand{\kmss}{km s$^{-1}$}
\newcommand{\km}{km s$^{-1}\;$}

\newcommand{\vlsr}{V$_{\rm LSR}$}

\newcommand{\vslsr}{V$_{\rm sys,LSR}$}

\newcommand{\lsun}{\mbox{L$_{\sun}$}}

\newcommand{\ho}{H$_{2}$O$\;$}

\newcommand{\mb}{mJy beam$^{-1}$}
\newcommand{\lwater}{L$_{\rm H_2O}\;$}

\newcommand{\hii}{H$_{\rm II}\;$}
\shorttitle{Low-Luminosity Extragalactic Water Masers in Active Galaxies}
\shortauthors{Yoshiaki Hagiwara}
\begin{document}

\title{Low-Luminosity Extragalactic Water Masers toward M82, M51, and NGC4051}

\author{Yoshiaki Hagiwara}
\affil{National Astronomical Observatory of Japan, Tokyo, Japan}
\email{yoshiaki.hagiwara@nao.ac.jp}
\begin{abstract}

Sub-arcsecond observations using the Very Large Array (VLA) are
presented of low-luminosity H$_2$O maser emission in M82,
M51, and NGC4051. New maser features have been detected within the
M82 starburst complex. They are largely associated with star-forming
activity such as optically-identified starburst-driven winds, H\,{\sc
ii} regions, or the early phase of star-formation in the galaxy.  \ho
maser in M51 consists of blue- and red-shifted features relative to
the systemic velocity of the galaxy.  The red-shifted features are
measured at the northwest of the nuclear radio source, while the
location of the blue-shifted counterpart is displaced by
$\sim$2" from the radio source.  A small velocity gradient
closely aligned with the radio jet is detected from the red-shifted
features. The red-shifted maser most likely amplifies the background
radio continuum jet, while the blue-shifted counterpart marks
off-nuclear star-formation in the galaxy.  All of the detected maser
features in the narrow-line Seyfert\,1 galaxy NGC4051 remain
unresolved by new VLA observations.  Due to the low luminosity of the
maser, the maser excitation is not directly related to the active
galactic nucleus. 
\end{abstract}
\section{INTRODUCTION}
Since the discovery of the first 22\,GHz extra-galactic \ho maser towards
\object{M33} by Churchwell et al.\ (1977), significant progress in
studies of extra-galactic \ho masers has been made, with
the first detection of a nuclear \ho maser toward an active
galactic nucleus (AGN) in NGC\,4945 \citep{doss79}.  A number of
single-dish surveys searching for new \ho masers in active galaxies
have been conducted since, in which more than 1000 galaxies were
observed.  These surveys were stimulated by the VLBI imaging of \ho maser
components in a sub-parsec--scale thin, warped, edge-on disk displaying 
Keplerian rotation around the nucleus of the Seyfert\,2 galaxy 
\object[M 106]{NGC 4258} \citep[e.g.,][]{herr98}. These surveys have
increased the total number of extra-galactic \ho masers to $\sim$80 at present
\citep{henk05,kond06}.

It is generally recognized that extra-galactic \ho masers can be grouped
according to their isotropic luminosity (L$_{\rm iso}$) \citep[e.g.,][]{linc93}. High-luminosity (L$_{\rm iso}$ $>$ 10 L$_{\odot}$) \ho masers are
associated with active nuclei.
Low-luminosity (L$_{\rm iso}$
$<$ 10 L$_{\odot}$) \ho masers are mostly associated with star-forming activity however, some of them may also contain or be in low radio luminosity AGNs.
The luminosities of Galactic \ho masers that are commonly observed in
the envelopes of evolved stars and star-forming regions range,
typically, from 0.001~L$_{\odot}$ to 1~L$_{\odot}$, and no
\ho maser with a luminosity well above 1~\lsun~ has ever been found
our galaxy.  Accordingly, \ho masers observed in nearby star-forming or starburst galaxies such as M33, M82, NGC253, and NGC6946 with 
L$_{\rm iso} values $ of 0.01--1~\lsun \citep[e.g.,][]{chur77, clau84, ho87, baud96}
are most likely to originate in star-forming activity in the host
galaxy.
On the other hand, there has been debate over the nature of a
sub-population of \ho masers with luminosities of 
1~\lsun~$\la$~L$_{\rm iso}$~$\la$~10~\lsun, most of which 
are considered to arise in
prominent sites of star-formation or starburst-activity in galaxies.
However, some of their host galaxies show AGN activity; low-luminosity 
water masers  have been detected towards LINER or type 1 Seyfert
nuclei in recent sensitive single-dish surveys
\citep{henk02,hagi03,braa04}.  They could simply be low-luminosity
analogues of the high-luminosity masers in narrow-line AGNs such as 
LINERs and type 2 Seyferts, and lower
maser luminosities might be explained by a close to face-on view of
obscuring structure around the nucleus, such as a disk or torus in
the line of sight, or a misalignment between the disk and a nuclear continuum
in the line of sight \citep{hagi03}.
It is important to explore the sub-population of extragalactic masers
with 1~\lsun~$\la$~L$_{\rm iso}$~$\la$10~\lsun~ at high angular resolution, 
as they
could advance the study of extra-galactic star-formation or probe the
nuclei in low radio luminosity AGNs. 

M82 (D = 3.63\,Mpc; \citet{free94}) is a well-known nearby
starburst galaxy with a long observing history at various
wavelengths. The galaxy hosts a number of exotic radio sources within
its starburst complex, including main- and satellite-line OH
masers and absorption, which trace cold molecular material in the galaxy
\citep[e.g.,][]{seaq97,mcdo02}.  Several low-luminosity \ho maser
features were detected at 1.4" resolution in earlier Very Large Array (VLA)
observations \citep{baud96} however, the resolution was not
sufficient to pin down or resolve the maser to investigate its
association with radio continuum sources such as the nucleus, jet,
compact H\,{\sc ii} regions, or supernova remnants (SNRs). 

The Whirlpool galaxy, M51 (D = 9.6 Mpc; \citet{sand75}), 
has provided an ideal laboratory for studying molecular materials in a galactic
star-forming environment. These materials are more abundant in the spiral arms than in the central region. The nuclear region is very 
complex, showing molecular emission that is asymmetric in both  position and
velocity with respect to the nucleus: the dominant red-shifted emission peak
is 1" to the west of the radio/optical nucleus 
\citep[e.g.,][]{scov83,scov98}. 
Ho et al.\ (1987) first reported the discovery of low-luminosity \ho 
maser emission towards the galaxy. A snapshot observation
pin-pointed the location of the known red-shifted maser features \citep{hagi01}
northwest of the continuum peak of the galaxy, while the 
blue-shifted feature(s) have not previously been measured.
Due to the insufficient velocity coverage of the earlier observation,
the overall distribution of the maser emission has not yet been explored in 
detail.

NGC4051(D = 9.7 Mpc; \citet{adam77}) is known to be a narrow-line
Seyfert\,1 galaxy (NLS1) showing `broad' emission line in H$\beta$ 
(FWHM\,$<$\,2000\,\kmss), narrower than that typically observed
(FWHM = 2000--10000\,\kmss) towards broad-line Seyfert\,1 galaxies \citep{oste85}. 
NLS1s are characterized by their housing of low-mass black holes 
radiating near their Eddington limits \citep[e.g.,][]{will04}.  NGC4051 
exhibits very strong X-ray intensity variability on many
different timescales \citep[e.g.,][]{lawr87}, which is interpreted as a
result of its nuclear activity and extended starburst components
dominated by soft X-ray emission \citep{sing99}.  \cite{hagi03}
detected an \ho maser in NGC4051 and identified it towards the center of the
galaxy with a complex nuclear radio structure \citep[e.g.,][]{ulve84}.
The maser luminosity is estimated to be $\sim$\,2 \lsun.  The maser
features span $\sim$300\,\kms and straddle the systemic velocity
of \vlsr = 730 $\pm$ 3 \kms \citep{hagi03} nearly symmetrically.
These Doppler-shifted maser features appear to indicate
the presence of a rotating disk however, there has been no compelling
evidence found to support the presence of a disk \citep{hagi03}. 
The origin of  low-luminosity \ho masers in NLS1s, including NGC\,4051,  remains unexplored.

In this paper, I report sub-arcsecond imaging of \ho masers
towards these three nearby galaxies, 
in order to update the earlier
observations with higher resolution and sensitivity and to further explore  the
nature of low-luminosity masers. All of them are so weak (flux
density $<$ 100\,mJy) that they cannot be studied in VLBI
observations.  Thus, sub-arcsecond imaging is currently the best method to
resolve the masers. In section 5 the statistical infra-red and
radio properties of the \ho masers are discussed.  Throughout this
paper H$_0$ = 75 \kmss Mpc$^{-1}$ is adopted.

\section{OBSERVATIONS}

Spectral line observations at  22 GHz using the NRAO\footnote{The National
Radio Astronomy Observatory (NRAO) is operated by Associated
Universities, Inc., under a cooperative agreement with the National
Science Foundation.}  VLA were carried out for
measurement of \ho maser emission (6$_{16}$-5$_{23}$ transition)
towards three nearby galaxies \object[NGC 3034]{M82}, \object[NGC
5194]{M 51}, and \object{NGC4051} from 2002$-$2003. All the observations
were performed on these galaxies when the array was in the A Configuration.
All the observations were made by employing two intermediate
frequency (IF) bands of width 12.5 or 6.25\,MHz with a 
single polarization, divided into 32 or 64 spectral 
channels, yielding velocity resolutions of 1.32 or 2.63 \kms 
per channel. Each IF band was centered on a velocity selected to cover all 
the maser emission, with respect to the local standard of rest (LSR).

Because low-luminosity extragalactic maser emission is difficult to detect
within an atmospheric coherence time at 22\,GHz, phase-referencing
observations were employed, using a nearby calibrator source
(Table~1) several degrees away from each galaxy. The
observations were thus performed by switching to the phase-referencing
source, typically every 2 minutes for 30 s.  Amplitude and
bandpass calibration were performed using observations of 3C286, and
the flux density scale was estimated to be accurate to within 10\%.
The data were calibrated and mapped in the standard way using the NRAO
Astronomical Image Processing Software (AIPS).  A summary of the
observations, including IF velocity ranges, synthesized beam
dimensions, and rms noise for each observing run, is provided in
Table~1.  
After the phase and amplitude calibrations, the image cube
was continuum subtracted using channels with no significant 
line emission in order to obtain only the line-emission.
Thus, the maser emission in M\,82 was separated out from the 
continuum emission. The continuum image of M\,82 was produced, 
collecting only line-free channels within a single 6.25 MHz IF band. 
In the $(u,v)$ data of M\,51 and NGC\,4051, no continuum emission 
was detected  to a 3 $\sigma$ rms noise level of 0.6 \mb ~for M\,51,  
and  0.48 \mb~for NGC\,4051. 

In this paper, the B1950.0 coordinate system is adopted for the
observations and figures of M82, since this system was used in many earlier works on M82 in 
the literature.
\section{RESULTS}
\subsection{M82}
\begin{figure}
\includegraphics[angle=0,scale=0.7]{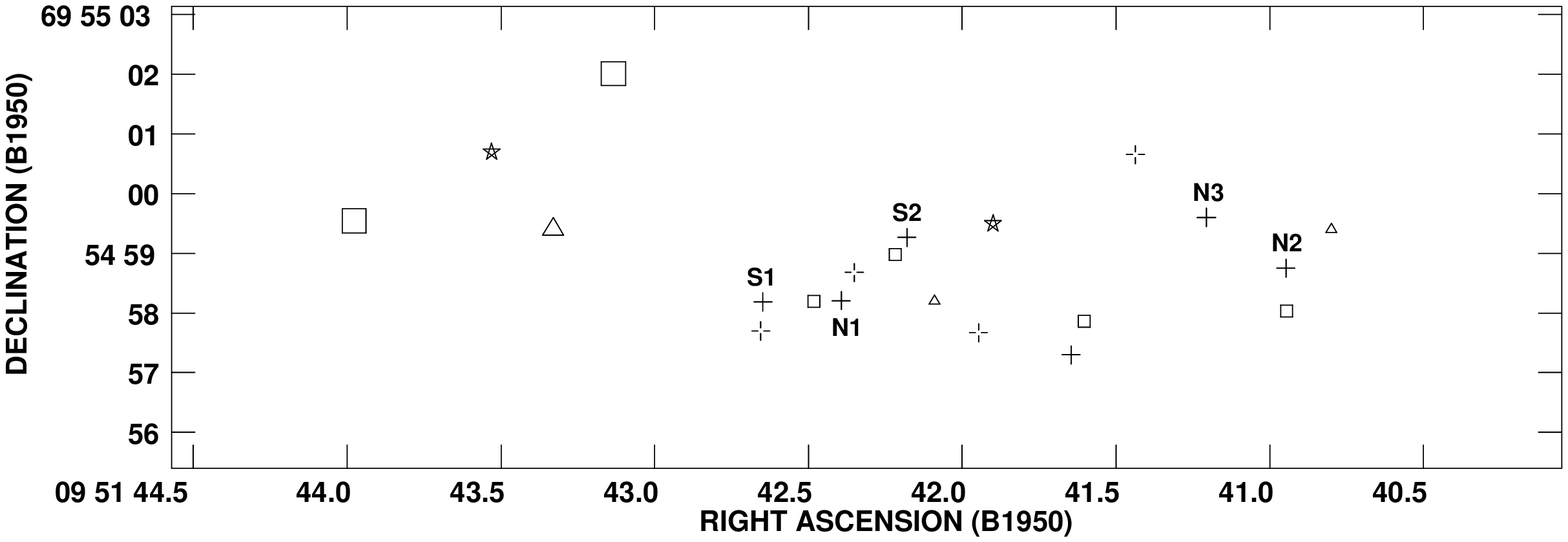}
\includegraphics[angle=-90,scale=0.7]{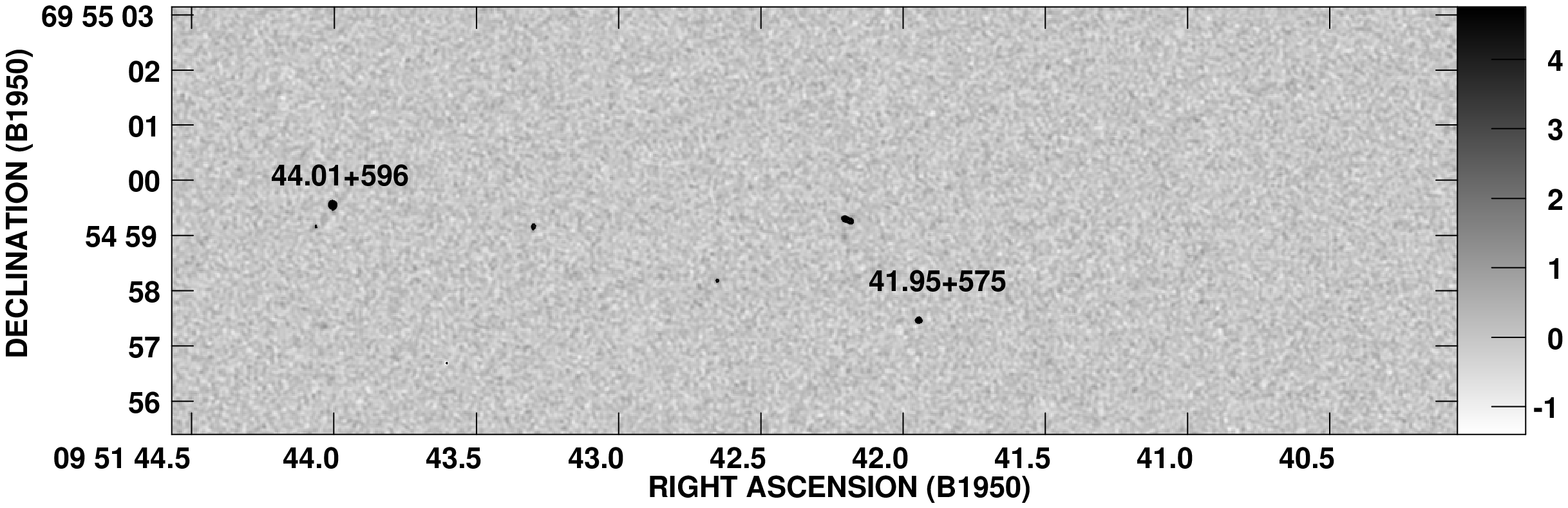}
\caption{$Top$: Locations of \ho masers, with other emission shown for
comparison, in M82.
Infra-red continuum peaks at 2.2 \micron
 ~\citep{diet86} are plotted with stars, \ho masers are by crosses, 1667/1665 MHz OH masers ~\citep{argo04} are shown by squares, 1720 MHz OH masers~\citep{seaq97} are shown by triangles, and  hard X-ray points are shown by broken crosses \citep{kaar01,mats01}.
 Larger squares and triangles
indicate blue-shifted emission relative to 200 \kmss; their smaller counterparts show red-shifted (\vlsr $>$ 200 \kmss) emission. The systemic velocity of the galaxy is 220 $\pm$ 5 \kmss. It is obvious that all of the locations of OH 
and \ho masers are well separated in space, depending on their offset relative
to the velocity of 200 \kmss. 
$Bottom$: The 1.3-cm radio continuum version of the map shown in the top panel.
Gray-scales are plotted from -1.0 to 5.0 mJy beam$^{-1}$.\label{fig1}}
\end{figure}

Figure~\ref{fig1} shows the locations of the detected \ho masers,
and other known sources in M82. 
A single-dish spectrum and individual VLA spectra of the \ho masers are
displayed in Figures 2 and 3.  All 
five VLA spectra had detections above the $\sim$5\,$\sigma$ level.
However, one might suggest that the detected features at N2 and N3 
need to be confirmed. All of these maser features remained unresolved at the 
angular resolution of $\sim$93\,mas, corresponding to 1.7\,pc at the adopted
distance, in the robust weighted maps.

The entire galaxy was searched, and no
other water maser emission was found in the velocity range of \vlsr = 50$-$150 \kmss. The known red-shifted emission \citep[e.g.,][]{clau84} in the galaxy
was not seen in this observation.  In addition to the four masers
S1--S4 (using the designation by \citet{baud96}), detected with  the VLA in
the C-configuration at 1.4" resolution \citep{baud96}, three maser sources
are newly detected in our observing run (Fig.~\ref{fig3}).  They are
hereafter labeled N1, N2, and N3, respectively. None of the positions
of the \ho masers coincide with those of  the OH masers, luminous X-ray sources, 
and infrared peaks that have been measured to date.

Table~\ref{tab:m82} shows the
positions, peak flux intensities, velocities, and velocity widths
 of the detected masers at the highest angular
resolution of 0.93". These new maser positions are constrained 
better than those at 1.4" resolution measured with the VLA in the C configuration by \citet{baud96}.  
The shapes of the detected maser
spectra that were reported in \citet{baud96} were relatively
unchanged since their observation in 1993.  The positions of S1 and S2
are consistent to within 0.08" between the two observations, and the
difference is most likely due to different interferometer
beam sizes.  No maser emission at or near the location of S3 and S4
was detected to a 3\,$\sigma$ level of 3.3\,\mb ~in this observation, 
probably due to intensity variation of the maser.
Table~\ref{tab:m82} lists the positions of the maser features at S2, in
which the emission profile is peaked at several velocities, spanning
$\approx$ 30 \kmss, but all lying in an $\sim$ 0.01"
region centered on R.A. (B1950)=09$^{\rm h}$51$^{\rm m}$42.1812$^{\rm
s}$, decl. (B1950)=69\degr54\arcmin59.260\arcsec.

Table~\ref{tab:m82cnt} lists the nearest discrete continuum source and 
radio recombination line (RRL) for each \ho maser.
In order to compare the positions between the masers and discrete
continuum sources in the literature \citep[][and references
therein]{mcdo02}, the published positions of two known sources at 15\,GHz 
were adjusted to those on our 22\,GHz continuum map, which led to a
sensible shift of $<$ 0.05" for 41.95+57.5 and 43.30+59.2.
Therefore, one can estimate that positions between the two
observations can be compared at an accuracy of $<$\,0.05".  The
maser positions in Table~\ref{tab:m82cnt} show an offset of 0.2--0.3"
(corresponding to 0.36--0.54\,pc) from the nearest continuum peaks. Such
a large separation makes it difficult to conclude that the detected
maser is associated with an H\,{\sc ii} region or SNR.  Thus, the
masers are not essentially associated with any continuum source at
this resolution. Alternatively, (compact) H\,{\sc ii} regions were not detected due to the sensitivity limit of these observations.
\subsection{M51}
Figure~\ref{fig4} displays VLA spectra of \ho masers detected
towards the central region of M51. The red-shifted features,
spanning \vlsr = 538 -- 586 \kmss, were previously detected in 
VLA snapshot observations \citep{hagi01}. 
One of the known blue-shifted features,
peaking at \vlsr\,=\,445\,\kmss, is detected by interferometry
for the first time. These Doppler-shifted velocity
features straddle the systemic velocity (\vslsr\,=\,472$\pm$3\,\kmss;
\citet{scov83}) of the galaxy.

Figure~\ref{fig5} shows the approximate locations of the blue- and red-shifted 
features on the 8.4\,GHz VLA-A image of \citet{brad04}. None of these features
is resolved by the uniformly weighted beam of 82\,$\times$\,77\,mas$^2$.
Assuming the masers have Gaussian distributions, 
the beam-deconvolved size of the red-shifted maser is $\sim$25\,mas,
corresponding to 1.2~pc at the accepted distance to the galaxy.
The positions of all the detected red-shifted features are within 
0.01$\arcsec$, centered on 
R.A. (J2000)=13$^{\rm h}$29$^{\rm m}$52$^{\rm s}$.709,
decl. (J2000)=+47\degr11\arcmin42\arcsec.79, 
which is consistent with 
the earlier observation by \citet{hagi01}. The blue-shifted feature
is located at 
R.A. (J2000.0)=13$^{\rm h}$29$^{\rm m}$52$^{\rm s}$.549,
decl. (J2000.0)=+47\degr11\arcmin43\arcsec.34. The parameters of the maser 
features are listed in Table~\ref{tab:m51}.

The projected distance between the blue- and red-shifted features
corresponds to 82\,pc (or 1.7$\arcsec$); the blue-shifted feature is well 
separated
from the other known features and so they are not kinematically related.
On the other hand, in both this and the earlier VLA observation, the red-shifted features are observed $\sim$180\,mas (or 8.6\,pc) northwest of the radio continuum peak of the galaxy \citep{hagi01}, suggesting that the maser does arise from the radio jet rather than in a low-luminosity nucleus of the galaxy.

Figure~\ref{fig5} also displays the first moment map produced
from the spectral-line data cube containing only 10 velocity channels
of red-shifted features in Fig~\ref{fig4}.  A weak velocity gradient,
roughly from south to north is seen along the axis P.A.\,=\,155\degr. 
The value of the gradient is, approximately, 
10\,\kms per 0.1", or 2\,\kms\,pc$^{-1}$.
The gradient is very weak but real, at least from 554 to 563\,\kmss, 
although the two extrema at the edges \vlsr\,=\,564 and 553\,\kms are
less certain. However, the zeroth-moment map is well sampled with a good
signal-to-noise ratio toward the center portions of the
source. As the VLA beam is 82\,$\times$\,77\,mas at P.A.\,=\,6$\degr$, 
there is no possibility that the gradient along P.A.\,=\,155\degr\,  
is arising from effects of the beam shape or orientation.
Accordingly, I conclude that the observed velocity gradient and 
its axis are real. 
\subsection{NGC\,4051}
VLA spectra of the \ho maser towards the center of NCC4051 (Fig.~\ref{fig6}) 
cover most of the known features in the range \vlsr\,=\,640--780\,\kmss.
The location of the systemic feature peaking at \vlsr\,=\,717.6\,\kms 
is measured
at R.A. (J2000)=12$^{\rm h}$03$^{\rm m}$09$^{\rm s}$.610,
decl. (J2000)=+44\degr31\arcmin52\arcsec.68, and all other 
features are confined to within about 0.01" (or 0.5\,pc)
from the systemic feature, resulting in the masers being unresolved
at this resolution.
The beam-deconvolved size of
each maser feature is smaller than 40\,mas, that is, about 2\,pc.
The result is consistent with the earlier VLA-A snapshot by \citet{hagi03}.
Although this new observation has improved ($u$,$v$) coverage 
over the previous single VLA snapshot, no new useful information is obtained.

\section{Discussion}
\subsection{M82}
\subsubsection{\ho Masers and Luminous X-Ray Sources}
The positions of the luminous hard X-ray sources in M\,82 were 
measured within an accuracy of 0.7" by {\it Chandra}
\citep{kaar01,mats01}; however, none coincide with those of the \ho masers
detected in this observation. 
In addition, there was  no high brightness radio source detected at 22\,GHz towards any of these luminous  X-ray sources, suggesting that these X-ray sources are not likely to
be low-luminosity AGNs. Since the masers do not overlap any
high-brightness radio sources like an AGN or jet, they are off-nuclear masers 
that are not directly amplifying the background continuum, as in the case of 
nuclear \ho masers. This is consistent with the fact that the 
maser luminosity in the galaxy is 1--5 orders of magnitude lower than those 
of the nuclear \ho masers. 
\subsubsection{\ho Masers and Other Sources}

None of the \ho masers detected in this observation are associated with
infra-red peaks at 2.2\,\micron, which were considered to be
low-luminosity AGN candidates or the dynamical center of the galaxy
\citep{diet86}, suggesting that the masers in M82 are not nuclear 
masers. The separations between each \ho maser peak and the
center of the nearest continuum source, given in Table~3, indicate that there
are no \ho masers at the locations of the H\,{\sc ii} regions or radio SNRs.  According to the 5\,GHz MERLIN and 15\,GHz
VLA-PT observations of continuum sources in the galaxy by
\citet{mcdo02}, the angular size of each  H\,{\sc ii} region appearing
in Table 3 is $\sim$ 0.1\arcsec--0.2\arcsec (2--4\,pc) in
radius. It is understood that the sizes of Galactic compact H\,{\sc ii} regions excited by central stars are typically 0.05--0.5\,pc and Galactic H\,{\sc ii} regions are less than $\sim$ 5 pc in radius \citep[e.g.,][]{gara99}.
Given the observed separations between the masers and the centers of
compact H\,{\sc ii} regions that range from 0.17\arcsec ~to 0.51\arcsec,
corresponding to 3--10\,pc, it is less plausible that
the masers in the galaxy are directly associated with H\,{\sc ii}
regions or ionizing central stars in the galaxy.  In contrast, VLBI observations revealed that the \ho maser in the nearby star-forming galaxy M33 is associated with the HII region IC133 \citep{linc93}. It is thus plausible that some other \ho features in M82 might be found in H\,{\sc ii} regions. The deconvolved angular sizes of RRL emission (H92$\alpha$) in the galaxy displayed in Table 3 range from 1.4\arcsec --1.7\arcsec
(25--30\,pc), just greater than the VLA synthesized beam of
0.9\arcsec~\citep{rodr04} that includes the masers at S1, S2, and N2.
The \ho maser would not have originated from the RRL -emitting ionized gas,  but the RRL marks the prominent star-forming regions in the galaxy, where maser excitation actively occurs. 

The velocity deviation of the ionized gas (100 -- 130\,\kmss) and
molecular gas (110 -- 145\,\kmss) from the galaxy's rigid
rotation velocity has been observed, and accounts for the
expansion of starburst-driven winds in the galaxy
\citep[e.g.,][]{mats05}, but the kinematics connecting the masering
clouds with the ionizing medium is not obviously seen from this observation.
\subsubsection{What Are the Masers in the Galaxy?}
The interpretation of the nature of the \ho masers in M82 is not
straightforward, since none have  distinct continuum
counterparts as effective pumping agents.  \citet{axel99} imaged an
expanding shell-like structure (superbubble) traced by CO and CO
isotopes in the galaxy, with a diameter of $\approx$130\,pc and
expansion velocity of $\sim$ 45 \kmss. According to their analysis, the
velocity of the approaching side of the expanding shell is 
\vlsr = 95\,\km and the receding velocity is \vlsr = 190\,\kmss. 
Thus, the former
has an \ho maser counterpart at \vlsr = 86  and 94 \kms, but the
latter is not covered in this observation. If the maser at S2 is
associated with the shell, it could be excited at the shock front of
the starburst-driven winds.  It is interesting to note that thermal
100\,GHz continuum emission has been imaged inside the CO superbubble,
and the distribution of the \ho masers roughly underlines the
continuum emission originating from starburst activity
\citep{mats05}. About 10 OH main-line masers are observed  that are
associated with blue-shifted parts of expanding wind-driven shells
\citep{axel99}, and the velocities and locations of these OH and \ho masers are 
largely consistent \citep{ap05,mats05}. These OH masers trace the 
1667 MHz OH absorption along the major axis of the galaxy,
and the OH dynamics understood as solid-body rotation agrees well 
with the CO emission \citep{shen95} and H\,{\sc i}  absorption \citep{will00}.
However, there are distinct deviations from this at 350 and $\sim$ 70 \kms
\citep{ap04}, although neither is  observed in our observations.
Therefore, the detected blue-shifted \ho masers are 
concentrated on the same molecular disk that is traced by 
OH absorption and CO emission; however, the association of the \ho masers
with the expanding shells or molecular outflows is not obvious
from our data.   

Similarly, in the starburst galaxy NGC\,253,  the locations of 
low-luminosity \ho maser ($\sim$ 1 \lsun) are measured towards 
the nuclear region with the VLA at $\sim$ 1" resolution \citep{henk04}.
 The association of the strongest blue-shifted maser features with the RRL, 
an expanding supernova, or starburst-driven winds is proposed , like the case in M82; however, there is no compelling evidence for that
from their observations. 

The extent of the CO(2--1) molecular clouds and outflows is not
inconsistent with that of the \ho maser  \citep{axel01}.  Molecular
outflows in our Galaxy are observed in \ho\ emission, typically on
scales of 1--10\,AU, which is at least 100 times smaller than the
masers measured on the parsec-scale with the VLA.  Accordingly, even if
the masers probed part of the outflow, they would not be resolved on
these scales by analogy with the cases of young stellar objects (YSOs)
in our Galaxy. It is believed that \ho masers indicate sites of
star-formation and appear at a certain stage of the evolutionary history
of protostars.  It is also understood that CH$_3$OH, OH, and \ho
masers are observed at different evolutionary stages of
star formation, and \ho emerges at the earliest stage of massive
star formation during the rapid accretion phase \citep{chur02}. 
Masers in a galaxy that do not accompany compact H\,{\sc ii} regions and 
OH masers may be the signposts of the early
stage of star formation and are most likely to be associated with
molecular outflows or accretion disks around extra-galactic YSOs that
have not been studied to date \citep[e.g.,][]{torr98}.
\subsection{M51}
\subsubsection{Blue-shifted Features}

The precise location of the known weak and
variable blue-shifted features lying from \vlsr= 435--445\,\kms in this 
galaxy \citep{hagi01} has been an open question.  One of them has been pinpointed in
this observation for the first time.  Taking the fact that the maser
luminosity of the blue-shifted emission is approximately 
0.17\,L$_{\odot}$ from the VLA spectrum and  is not associated
with any known radio continuum sources in the galaxy, one can infer
that AGN activity is not the major source giving rise to the maser.  The
blue-shifted emission is displaced 1.7", or 82\,pc, from the
red-shifted counterpart near the nucleus, thus, no hint of physical
connection exists between these velocity clusters.  From what do the
features arise?  The peak velocity of \vlsr = 445\,\kms is in the
velocity range of the central HCN emission \citep{kohn96}.  The
 features lie in the central dense region of the HCN
emission, where the mean velocity field of the HCN emission is \vlsr
$\sim$ 480\,\kms, that is,  35\,\kms shifted from the maser
\citep{kohn96}.  Thus, it is impossible to find common mechanics
connecting the two different velocity gradients traced by the maser
and the HCN. All that one can speculate is that the maser traces the
most dense part ($\ga$\,10$^7$\,cm$^{-3}$) of the circumnuclear
molecular gas within 10\,pc of the nucleus.  As a result of this
observation, the bipolar-jet model proposed in \cite{hagi01}, in which
the blue-shifted features are associated with the approaching side of
the jet, is no longer eligible.

\subsubsection{Central Kinematics}
One of the most exciting results from this observation is that part of
the velocity structure of the dense molecular gas within a few parsecs
of the nucleus is resolved at sub-arcsecond resolution.  The
detected velocity gradient is about 2\,\kms\,pc$^{-1}$, and the
direction of the gradient (PA\,$\sim$\,155\degr) is similar to that of
the radio jet but not to the rotational axis of the inner torus 
with a radius
of 70\,pc (PA $\sim$ 160\degr --165\degr) postulated from the
distribution of the HCN (J = 1$-$0) emission observed at 4" resolution \citep{kohn96}. 
Most of the red-shifted maser features in
Fig.~\ref{fig4} are neither in the velocity range of CO (J = 1$-$0)
(\vlsr = 387--547\,\kmss) nor in that of HCN (J = 1$-$0) (\vlsr =
377--546\,\kmss). The velocity ranges of the maser do not agree with 
those of the HCN emission and  the HCN torus model proposed in \citet{kohn96},
however, the axis of the HCN torus is not inconsistent with the 
direction of the maser velocity field.  

Given the fact that the blue-shifted features are separated 
by $\sim$80\,pc from the rest of the features, and do not coincide with the
nucleus, the whole maser system is unlikely to trace a
subparsec-scale Keplerian disk observed with the nuclear \ho maser, if it exists.
However, it is possible that only the red-shifted features trace a
part of a rotating disk, and the blue-shifted features on the disk are
invisible. Assuming that the observed maser is on a disk at a radius $r$ from the nucleus and 
with a rotating velocity of V$_{rot}$, the
observed velocity gradient (d$V$/d$l$) along P.A. = 155$\degr$ can be
expressed as d$V$/d$l$ $\simeq$
d(V$_{rot}$~$\theta$~sin$i$)/d($r$$\theta$) = $V_{rot}$/$r$, where $l$
is the projected distance, $i$ is the disk inclination, and
the approximation of sin$\theta \approx \theta$ is made. Adopting the disk
inclination of 80$\degr$ from the edge-on circumnuclear-nuclear disk
with a radius of 70\,pc \citep{kohn96}, together with  the observed
parameters of $dV/dl$ = 2\,\kms\,pc$^{-1}$ and $V_{red} - V_{sys}$ =
$V_{rot}$ $\simeq$ 100 \kmss, $r$ is estimated to be 50\,pc. The result
suggests that the observed maser probes a disk on a scale of 10\,pc but
not a thin disk with $r <$\,0.1\,pc, such as the one in NGC\,4258
\citep[e.g.][]{herr98}.

Does the maser disk lie in an inner part of the circumnuclear HCN
disk?  Taking this value for the radius of a molecular disk, the mass
confined within the disk is 1.2$\,\times\,10^8\,M_{\sun}$, that
is, 1 or 2 mag larger than the molecular hydrogen and
dynamical mass within $r$ $<$\,70\,pc, based on the estimation from the
HCN intensities \citep{kohn96}. Accordingly, we cannot account for the
detected velocity field with a subparsec-scale masering disk or a larger
scale molecular disk.

It is interesting to note that the velocity gradient of 
0.65\,\kms\,pc$^{-1}$ along the radio jet axis was detected 
in CO(J = 3--2) at 4\arcsec resolution ($\sim$160\,pc) 
using the Submillimeter Array by \citet{mats04}, although the sampled area in their
image is more than 100 times larger than that in this VLA
observation. The CO velocity gradient in their position-velocity map
covers that of the maser at \vlsr = 555--565\,\kmss. However the
observed CO gradient is more dominant in the direction
perpendicular to the jet, and such a small velocity gradient in the
thermally excited molecular gas is likely due to the internal
turbulence, as also mentioned  in \citet{mats04}.  Thus, there is no
compelling evidence that the maser and CO(J = 3--2) trace the same
kinematics, but both of them could be distributed in the same molecular cloud. In the VLBI observation of the 1667\,MHz OH maser
towards the type~1 Seyfert nucleus in Mrk\,231, a weak velocity
gradient of 1.4\,\kms\,pc$^{-1}$ across the torus structure with a
radius of 65\,pc is reported in OH emission \citep{hans03}, which does
not agree with any thermal molecular gas structure. In this sense,
the OH maser in Mrk\,231 is similar to the \ho maser in M51.
The axis of the OH torus in Mrk\,231 is not aligned with the major axis of the nuclear continuum source; rather,  it is perpendicular to the nuclear continuum axis. This is different from the case of M51, in which the velocity field of the maser is almost aligned with the jet axis, which might rule out the presence of a masering disk in the galaxy.
Note that there are still some other blue-shifted features whose positions have not yet been pinned down,  some of which could be on a masering torus 
within 1 pc from the center, along with red-shifted counterparts. If 
the maser in M51 is resolved down to scales of 0.1\,pc using VLBI, the velocity structures of the maser will be interpreted differently.
The most plausible explanation for the maser and its velocity gradient
based on these observations is that the maser arises in a dense molecular environment in the foreground of a  radio jet, amplifying the background radio continuum jet \citep{clau98},  or, alternatively is from a shocked dense 
region near the boundary of the jet \citep{hagi01}.
The weak radio intensity of the jet would account for
the low luminosity of the maser.

\subsection{NGC4051}

The luminosity of the \ho maser in NGC\,4051 is $\sim$1\,L$_{\odot}$, very
low for nuclear masers. It might appear that such a low-luminosity
maser is not related to direct AGN activity.  However, \citet{hagi03}
proposed that the low-luminosity of the maser and narrower total
velocity span of the Doppler-shifted features ($\sim$100 \kmss) might
be due to low-gain maser amplification, resulting from a small
inclination of a disk-like structure surrounding the AGN.
\citet{hagi03} hypothesized that the maser originated from a
less edge-on disk-like configuration surrounding a nucleus, which would cause
relatively short gain paths in the line of sight.
This could account for the low-luminosity of the maser.
\citet{made06}, however, argue that the maser in NGC\,4051 is unlikely to be
connected with direct AGN activity, in that the X-ray absorbing medium
is significantly ionized in NGC\,4051, and such an ionized absorbing medium
is unlikely to be physically close to the masering
medium.  They suggest that
the origin of the maser in NGC\,4051 is in nuclear winds, because the broadly 
spread ($\sim$ 280 \kmss) narrow -line spectrum is similar to that 
of the Circinus galaxy. Circinus hosts
highly Doppler-shifted maser features with a total velocity range of
$\sim$160\,\kms that are introduced as nuclear wind components,
together with features tracing an edge-on disk with a
sub-Keplerian rotation \citep{linc03}. However, a physical
mechanism driving the nuclear winds in the galaxy has not yet been proposed.\\

One can speculate that the medium giving rise to the maser is
geometrically separated from the X-ray absorber. The inner edge of the
absorber could be illuminated by direct X-ray radiation and hence
ionized, which would also cause a high time variability of the X-ray-measured column density $N_H$, while the masering medium could exist in the outer edge of a thin disk
that is much less ionized because of a large radial distance from the
X-ray heating source.  The known periodic intensity variability at
hard X-ray bands might occur in the inner edge of the disk
\citep[e.g.,][]{herr98}, while the time-interval of the intensity
variability of the maser is very different. In the case of NGC\,4388, 
the radius of the absorbing medium is estimated from the variability of 
$N_H$ to be a few hundred Schwarzschild radii ($R_{\rm {sch}}$) \citep{elvi04}, that is 100 times smaller than the (sub-)parsec-scale obscuring medium observed
in the \ho maser. \ho masers in NGC\,4258 are distributed in a region between 
 40,000 and 80,000 \,$R_{\rm {sch}}$ from the nucleus \citep[e.g.,][]{herr98}. Therefore, it is less plausible that the masering medium and X-ray-ionizing structure
occur in a single physical structure.  Thus, we still cannot  rule
out the presence of a maser disk in NGC\,4051.
In any case, the hypothesis proposed by \citet{hagi03}
should be tested at milliarcsecond angular resolutions by resolving
the maser distribution on scales of 0.1\,pc or less.
\section{FAR-INFRARED VS RADIO INTENSITY CORRELATION DIAGRAM}

There is a known strong linear correlation between the far-infrared
(FIR) flux density and the radio flux density at 1.4\,GHz from nuclear
starburst galaxies \citep{helo85,cond91}, which implies that FIR radiation
and non-thermal radio emission at 1.4\,GHz are closely connected with
star-forming activity.  In Fig~\ref{fig7} there are three 
different comparisons: the ratios of the 1.4\,GHz radio fluxes to FIR flux
from high-luminosity \ho masers, as well as from low-luminosity \ho masers
and prototypical OH megamasers (OHMMs) from ultra-luminous FIR galaxies.
 The low-luminosity masers and OHMM samples follow well
the known FIR-radio correlation.
The mean constant ratios of these three samples are also 
displayed  in Fig~\ref{fig7}, from which one can see that the ratio
for the low-luminosity masers is steeper than those of the other two.
The reason for this deviation is that the FIR fluxes
are relatively more dominant for the low-luminosity masers, which 
supports the idea that the low-luminosity masers are powered by star-forming
activity rather than AGN activity.  
We cannot discriminate the synchrotron emission at 1.4 GHz from either SNR or AGN ejecta, 
but the FIR emission is characterized as re-emission from the dust heated by starburst/star-forming  activity \citep[e.g.,][]{hagi02}.
It is interesting to note that there is a population of high-luminosity masers that
shows a higher ratio of the radio to FIR fluxes in the bottom phase-space in the diagram, which may suggest a close connection of 
maser excitation to  strong radio flux from AGN activity.
\section{SUMMARY}
Low-luminosity extragalactic \ho\ masers have been identified in
different physical environments related to star-formation, such as
H\,{\sc ii} regions, YSOs, or possibly SNRs. Given the fact that the
luminosity of the strongest Galactic \ho maser, W49N, is 
$\sim$\,1\,L$_{\odot}$, the origin of these low-luminosity masers can be
explained by star-forming activity in their host galaxies.  Many of
the host galaxies of these masers contain star-forming regions 
or exhibit starburst activity but do not contain an AGN.
These \ho\ masers may reveal new stellar phenomena relevant to 
extragalactic star-forming activity. 

The kinematics of \ho masers in M82 is broadly consistent with
OH or CO solid body rotation along the galaxy's  major axis, 
but locally the masers are not associated with any other molecular emission.
Including the three new detections, the masers are not
directly associated with any continuum sources , such as compact 
H\,{\sc ii} regions or radio SNRs, suggesting that the masers arise from the
earliest stage of star-forming activity in the galaxy. For the masers
in M51, the detected blue-shifted features are significantly offset
from the nuclear radio continuum, while the red-shifted counterparts
are in the nuclear radio source. Seemingly, the maser in the galaxy
amplifies the radio jet continuum, which needs to be confirmed at
higher resolution. The most remarkable result is that a small velocity
gradient roughly along the jet has been detected from red-shifted
emission at 0.1" resolution. However, the interpretation of 
the velocity gradient is not straightforward.
The nature of the maser in NGC4051 is controversial and also needs to be studied at higher resolutions. A study of the distribution of the rare \ho masers towards a NLS1 showing 
a low accretion rate and hence a smaller black hole mass is of great interest.
According to a statistical study, it is found that the mean FIR-to-radio flux ratio of  low-luminosity \ho masers is higher than that of their high-luminosity counterparts, suggesting that the maser excitation of low-luminosity water masers such as the one in M82 is  related primarily to star-forming activity, rather than AGN activity. 
\acknowledgments
I would like to thank the anonymous referee for a number of useful comments on 
the manuscript.
I am grateful to the NRAO staff for their assistance during data
analysis. This research is based on observations with the 100m telescope of 
the Max-Planck-Institut f$\ddot{\rm u}$r Radioastronomie at Effelsberg.
This research has made use of the NASA/IPAC Extragalactic Database,
which is operated by the Jet Propulsion Laboratory, California
Institute of Technology, under contract with the National Aeronautics
and Space Administration.

\clearpage

\include{tab1}
\include{tab2}
\include{tab3}
\include{tab4}
\onecolumn
\begin{figure}
\epsscale{.80}
\plotone{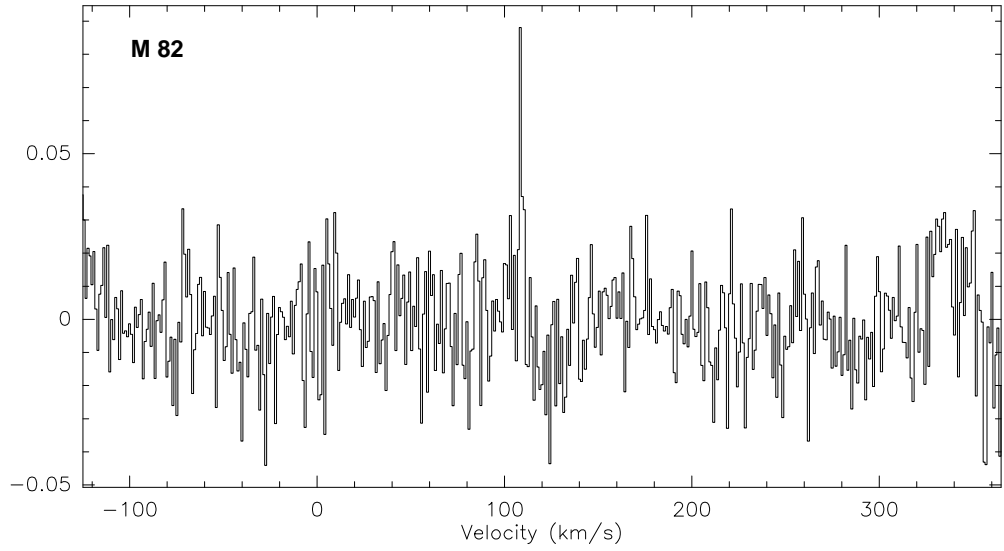}
\caption{Single-dish spectrum of the \ho maser towards the center of
  M82. The velocity resolution is 1.1 \kmss. The total integration
  time is $\sim $ 60 minutes, observed with the MPIfR 100 m telescope
  at Effelsberg on 7 March 2002.  The amplitude is scaled in janskys.\label{fig2}}
\end{figure}
\begin{figure}
\epsscale{.75}
\plotone{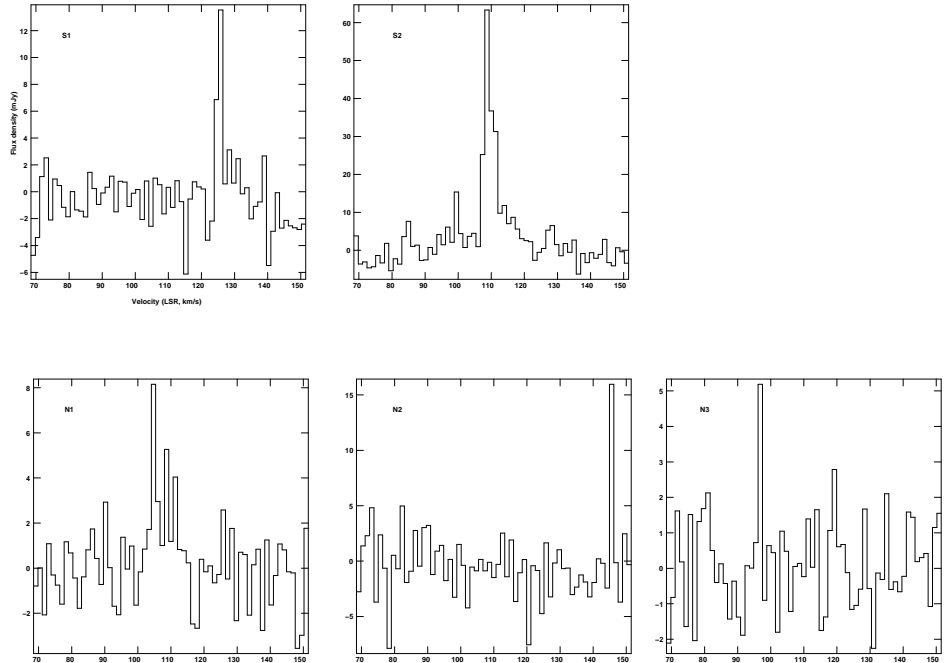}
\caption{
VLA \ho maser spectra towards five locations in M82 within \vlsr =
70 --150 \kmss. The masers towards S1 and S2 were originally
detected at 1" resolution of the VLA \citep{baud96}. Their
positions are constrained at 0.1" resolution in this paper. The
masers at N1, N2, and N3 are new detections.
\label{fig3}}

\end{figure}
\begin{figure}
\epsscale{.8}
\plottwo{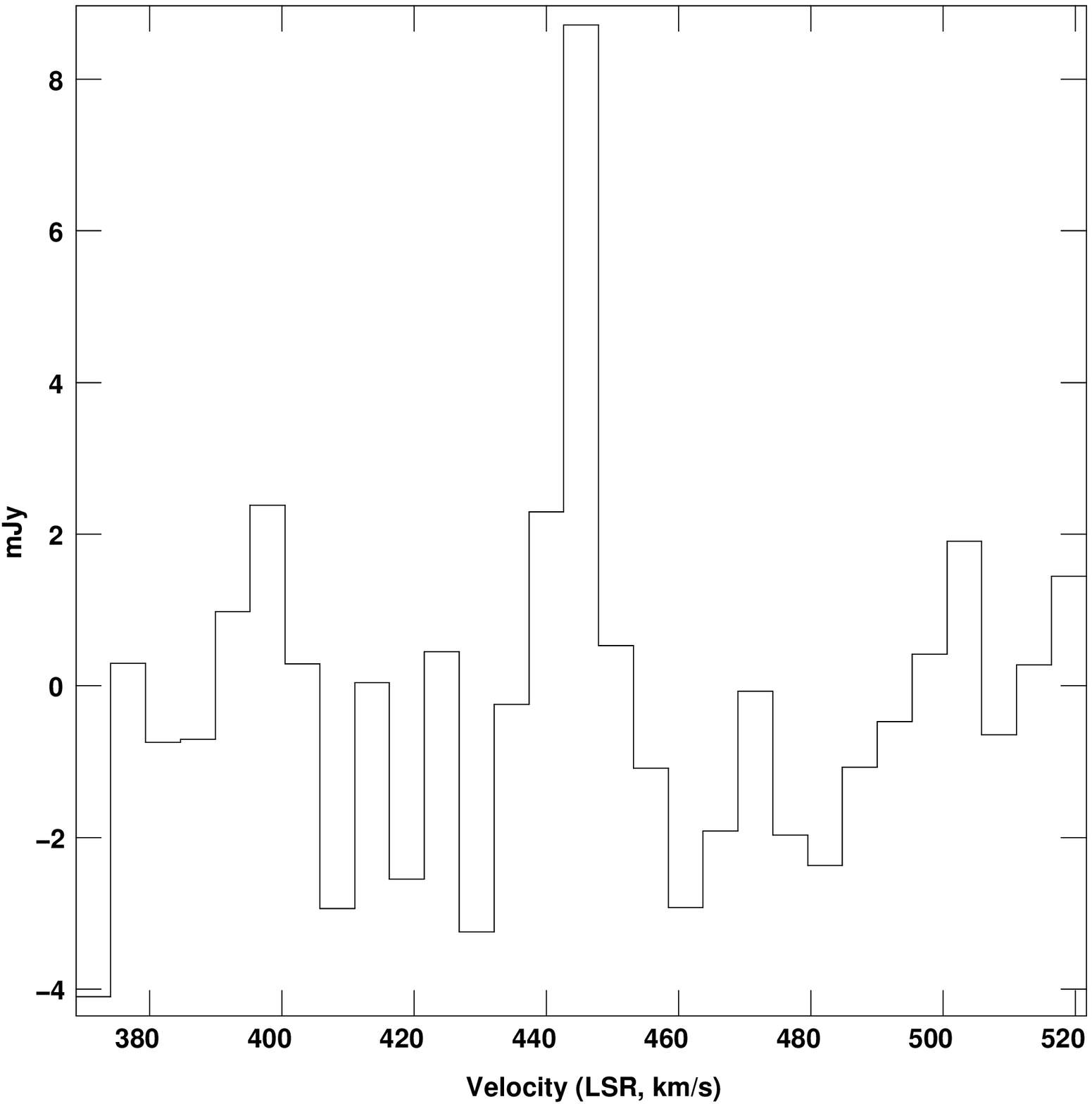}{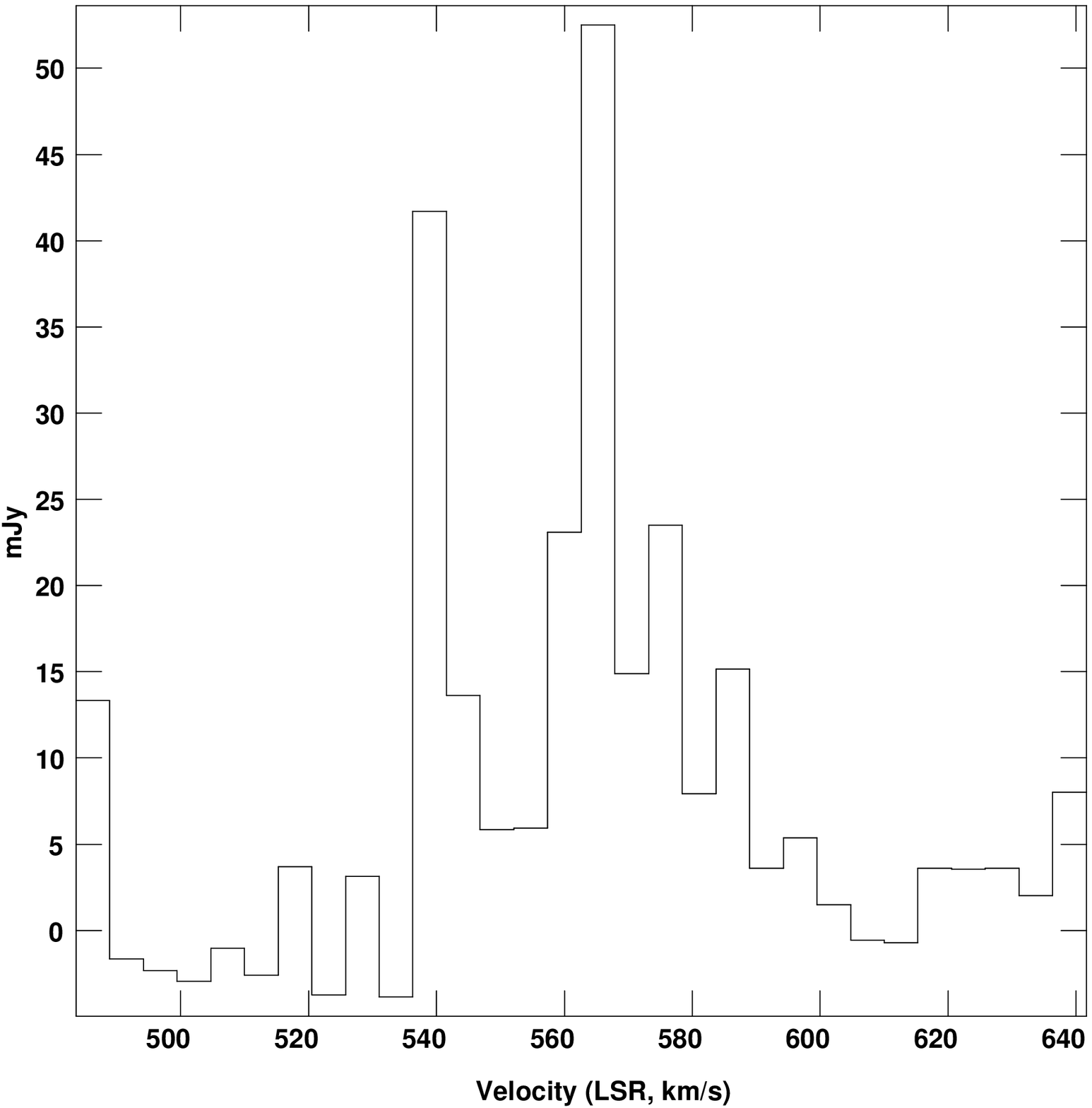}
\caption{\ho maser spectra of M51, observed with VLA-A on 4 July 2003. The
left panel covers blue-shifted velocity range, and the right panel covers the
red-shifted range. The velocity resolution is 5.3\,\kmss.  The spectra
were obtained from two IF channels, each with 12.5\,MHz bandwidth.
\label{fig4}}
 \end{figure}
\begin{figure}
\epsscale{.9}
\plotone{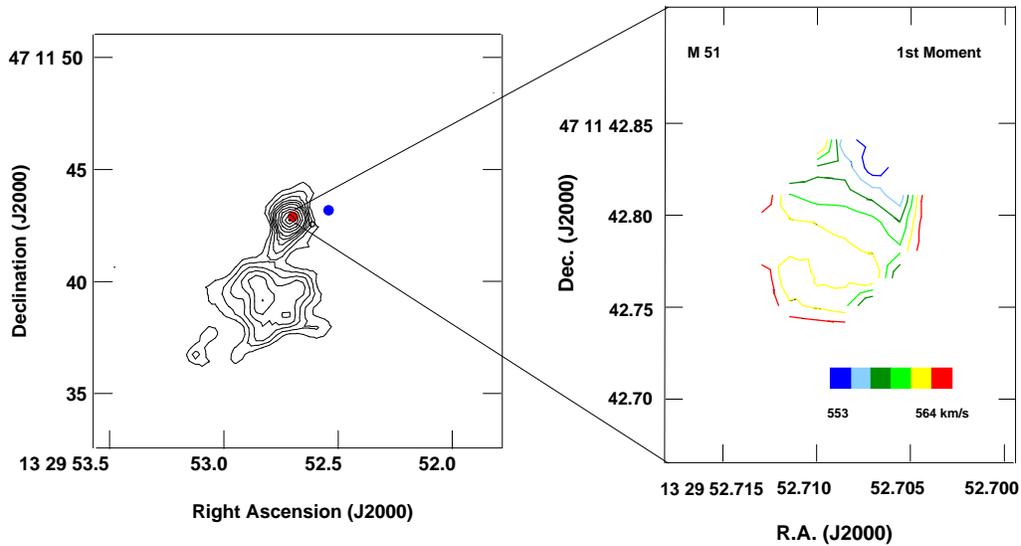}
\caption{
$Left$: Large-scale 8.4 GHz nuclear continuum of M51 imaged by the VLA
\citep{brad04}, superposed with two approximate positions of the \ho maser.  
The red-shifted cluster is marked in red in the vicinity of the continuum
peak and the blue-shifted feature(s)  marked in blue is offset from both the
peak and the other maser cluster. $Right$: First velocity moment map
of the red-shifted maser from the uniform-weight map. A small velocity
gradient over the maser is identified. Contours are spaced at 1--2 \kms
and labeled with the absolute LSR velocity-scale.  \label{fig5}}
\end{figure}

\begin{figure}
\plottwo{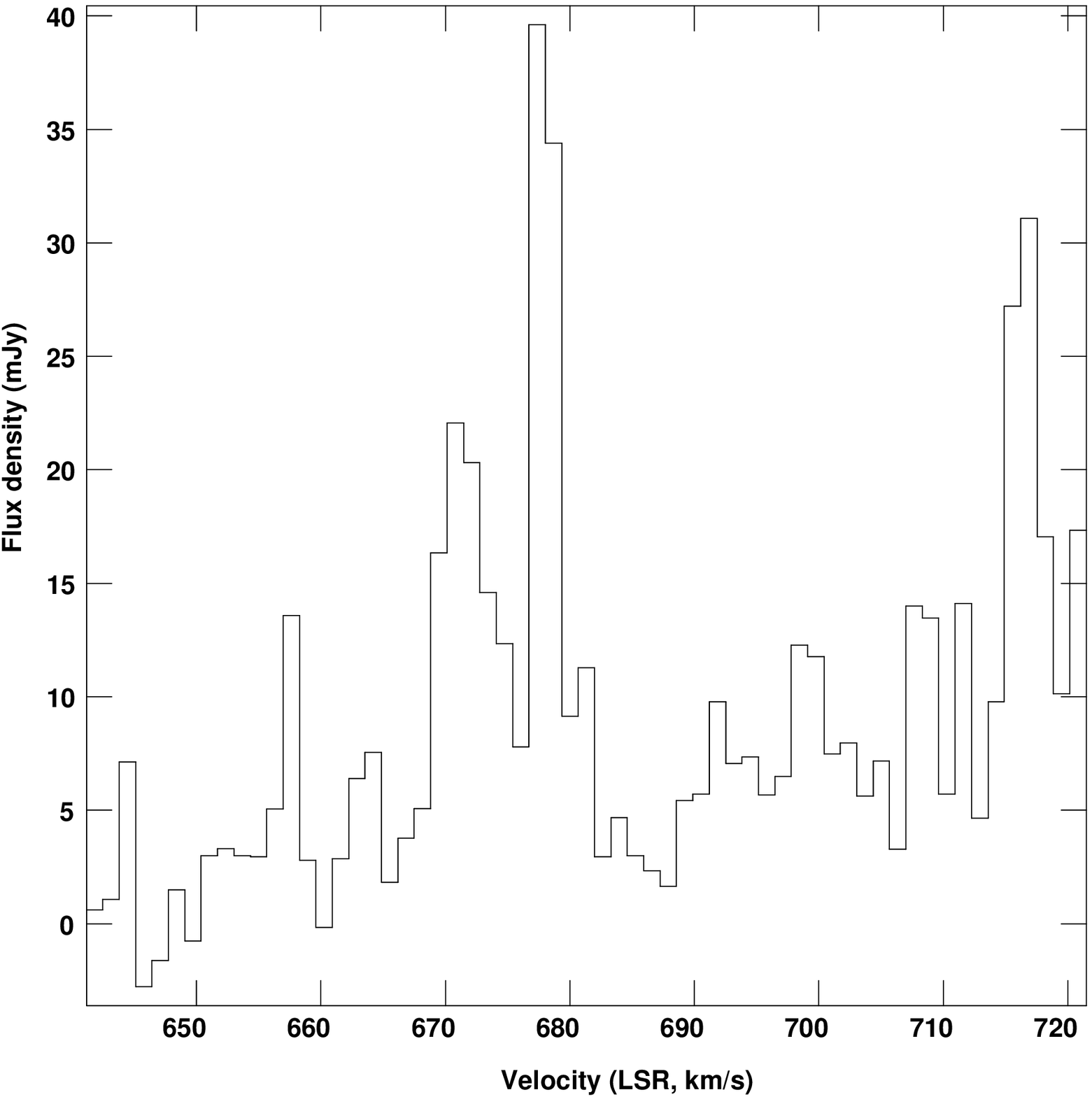}{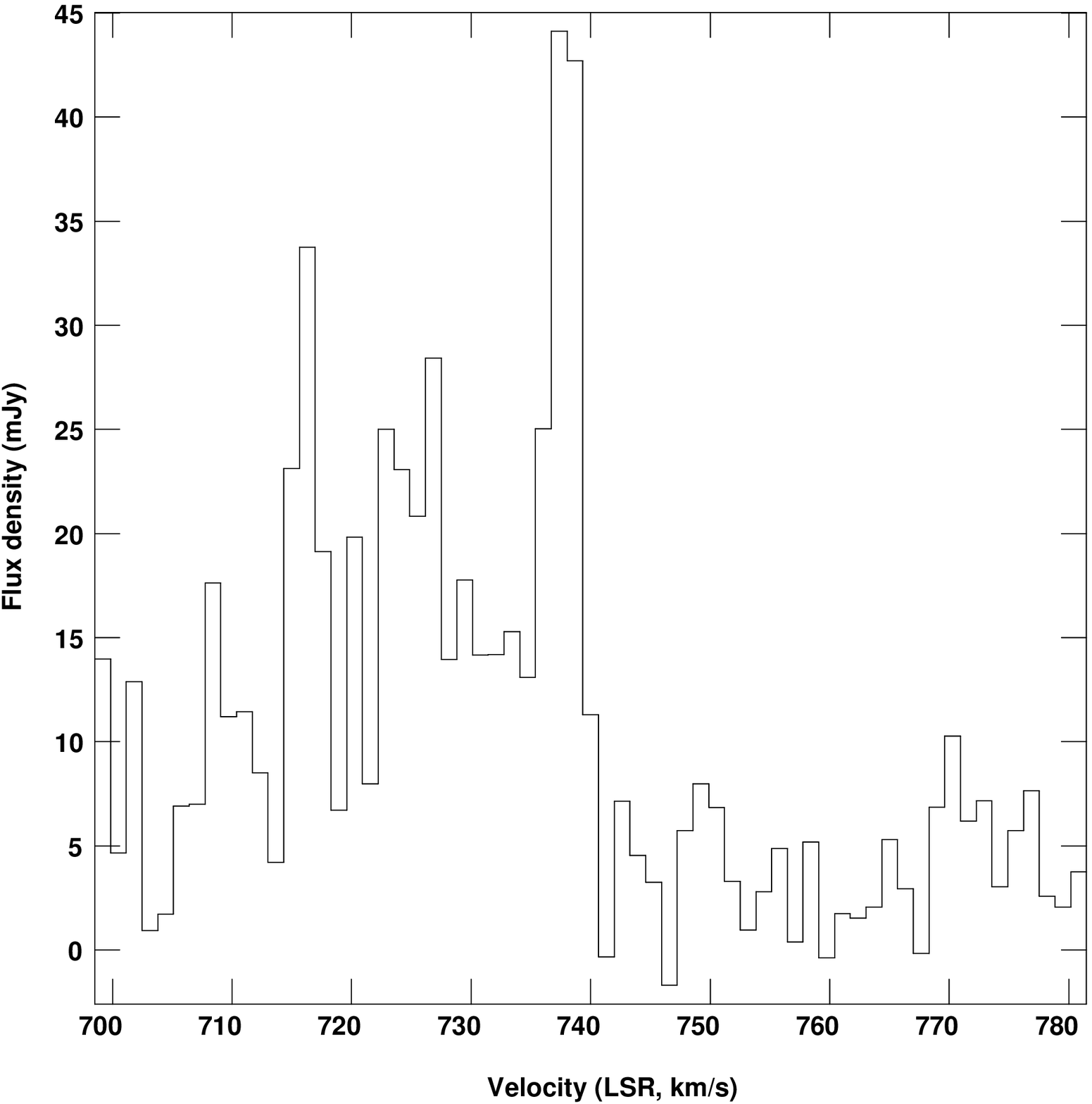}
\caption{\ho maser spectra from the nucleus of NGC\,4051 with a velocity 
resolution of 1.3\,\kms using the VLA. 
The spectra consist of two independent IF channels that cover
the velocity range of \vlsr = 640$-$780 \kms, straddling the systemic velocity
of \vlsr= 730\,\kmss. \label{fig6}}
\end{figure}
\begin{figure}
\epsscale{0.8}
\plotone{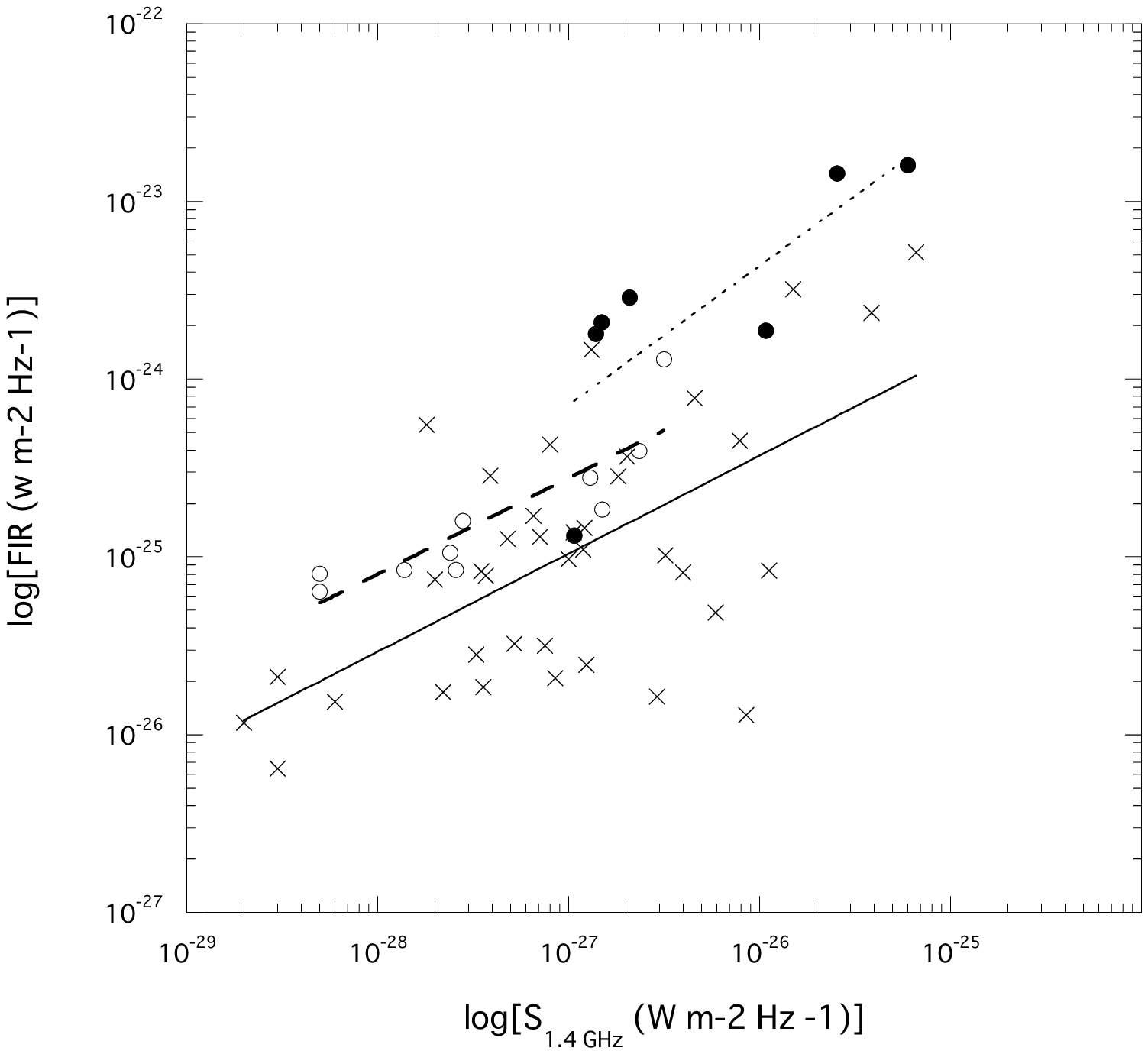}
\caption{The 1.4\,GHz radio flux density vs. FIR flux  density (estimated
from 60\,$\micron$ and 100\,$\micron$). Crosses represent the known
high-luminosity water masers in the literature, and filled circles represent
low-luminosity water masers in star-forming nuclei (IC10, IC342,
M51, M82, NGC253, NGC2146, and NGC6946), primarily from
\citet{linc90,tar02a,tar02b, henk02,clau84}. Open circles represent
prototypical OHMMs from Sanders et al.\ (1988). Mean
constant ratios of FIR to 1.4 GHz radio flux densities for
the three different samples are indicated by the solid line (high-luminosity
\ho maser), the dashed line (OHMM), and the dotted  line (low-luminosity \ho
maser).
\label{fig7}}
\end{figure}

\end{document}

%% file: tab1.tex
\begin{deluxetable}{rrrrrrrrr}
\tabletypesize{\scriptsize}
\tablecolumns{9}
\tablewidth{0pc}
\tablecaption{Summary of VLA Observations}
\tablehead{
\colhead{Object}&\colhead{Date}&\multicolumn{2}{c}{Tracking-Center} 
& \colhead{Velocity}&\colhead{$\Delta$V$^a$}&
\colhead{t$^b$}&\colhead{Beam}&\colhead{$\sigma_{\rm rms}$} \\
\cline{3-4} 
\colhead{\it (Cal. Source)}& \colhead{}& \colhead{R.A.}& 
\colhead{Decl.} & 
\colhead{Range} & \colhead{}& \colhead{} & \colhead{(HPBW)} & \colhead{(mJy}  \\
\colhead{} & \colhead{(d.m.y)}& \colhead{(h m s)}& \colhead{(\degr~\arcmin~\arcsec)} & 
\colhead{(\kmss)} & \colhead{(\kmss)} & \colhead{(hr)} & 
\colhead{(mas)}& \colhead{beam$^{-1}$)}  
}
\startdata
 &&\colhead{$\alpha_{1950}$}&\colhead{$\delta_{1950}$}&&&&& \\
\colhead{M82}&\colhead{13.4.02}&\colhead{09 51 42.200}&\colhead{69 54 59.30}&\colhead{50-130}&\colhead{1.3}&\colhead{9}&\colhead{93$\times$80}&\colhead{1.1} \\
    &&&&\colhead{70-150}&&&\colhead{PA=45\degr}& \\
\colhead{\it (0955+654)} &\colhead{-}&\colhead{09 54 57.847}&\colhead{65 48 15.53}&\colhead{-}&\colhead{-}&\colhead{-}&\colhead{-}&\colhead{-}  \\
\hline 
 &&\colhead{$\alpha_{2000}$}&\colhead{$\delta_{2000}$}&&&& \\
\colhead{M51}&\colhead{4.7.03}&\colhead{13 29 52.708}&\colhead{47 11 42.79}&\colhead{365-525}&\colhead{5.3}&\colhead{8}&\colhead{82$\times$77}&\colhead{1.1} \\
    &&&&\colhead{480-640}&&&\colhead{PA=6\degr}& \\
\colhead{ \it (1419+542)}&\colhead{-}  &\colhead{14 19 46.597}&\colhead{54 23 14.78}&\colhead{-}  &\colhead{-}&\colhead{-} & \colhead{-} &\colhead{-}  \\
\hline 
 &&\colhead{$\alpha_{2000}$}&\colhead{$\delta_{2000}$}&&&& \\
\colhead{NGC4051}&\colhead{4.7.03}&\colhead{12 03 09.606}&\colhead{44 31 52.52}&\colhead{640-725}&\colhead{1.3}&\colhead{9}&\colhead{77$\times$72}&\colhead{1.3} \\
  &&&&\colhead{695-780}&&&\colhead{PA=81\degr}& \\
\colhead{\it (1153+493)}&\colhead{-}&\colhead{11 53 24.466}&\colhead{49 31 08.83}&\colhead{-}&\colhead{-}&\colhead{-}&\colhead{-}&\colhead{-} \\
\enddata
\tablenotetext{*}{Equinox  for M82 is B1950.0; equinox for M51 and NGC4051 is J2000.0.}
\tablenotetext{a}{Velocity resolution}
\tablenotetext{b}{Total observing time}
\end{deluxetable}

%% file: tab2.tex
\begin{deluxetable}{crrcccr}
\tabletypesize{\scriptsize} \tablecolumns{7} \tablewidth{0pc}
\tablecaption{\ho maser in M82\label{tab:m82}} \tablehead{
\colhead{Component}&\colhead{$\alpha_{B1950}$}&\colhead{$\delta_{B1950}$}&\colhead{Peak Flux}&\colhead{Velocity}&\colhead{Feature Width\tablenotemark{a}}&
\colhead{\lwater}\\
\colhead{}&\colhead{(09$^{\rm h}$51$^{\rm m}$00$^{\rm s}$)}&
\colhead{(69\degr54\arcmin00\arcsec)}& (mJy)&\colhead{(\kmss,LSR)}&\colhead{(Channel)}&
\colhead{(\lsun)} } \startdata
\colhead{S1\tablenotemark{b}}&\colhead{42.6476 $\pm$
0.0006}&\colhead{58.180 $\pm$ 0.003}& 13&126&2&0.008\phn \\
\colhead{S2\tablenotemark{b}}&\colhead{42.1808 $\pm$ 0.0002}&\colhead{59.263 $\pm$
0.001}&63&108&3&0.062\phn \\ 
\colhead{}&\colhead{42.1799 $\pm$
0.0007}&\colhead{59.257 $\pm$ 0.004}&15&98&1&0.006\phn \\
\colhead{}&\colhead{42.1825 $\pm$ 0.0014}&\colhead{59.252 $\pm$
0.007}&8&84&1&0.003\phn \\ 
\colhead{N1} &\colhead{42.3951 $\pm$
0.0012}&\colhead{58.195 $\pm$ 0.007}&8&105&4&0.007\phn \\ 
\colhead{N2}
&\colhead{40.9436 $\pm$ 0.0007}&\colhead{58.756 $\pm$
0.004}&16&146&1&0.006\phn \\ 
\colhead{N3} &\colhead{41.2053 $\pm$
0.0014}&\colhead{59.602 $\pm$ 0.007}&5&98&1&0.002\phn \\ 
\enddata
\tablenotetext{a}{Each channel is 1.3 \kms wide.}
\tablenotetext{b}{Adopting labels in \citet{baud96}}
%
\end{deluxetable}

%% file: tab3.tex
\begin{deluxetable}{ccccccc}
\tabletypesize{\scriptsize}
\tablecolumns{7}
\tablewidth{0pc}
\tablecaption{\ho Maser and Other Sources in M82\label{tab:m82cnt}}
\tablehead{\colhead{}&\multicolumn{3}{c}{Nearest Discrete Continuum $^a$}&\colhead{}&\multicolumn{2}{c}{Nearest H92$_{\alpha}$$^b$ RRL}\\
\cline{2-4}\cline{6-7}
\colhead{\ho Maser}&\colhead{$\alpha_{B1950.0}$}& \colhead{$\delta_{B1950.0}$}&\colhead{$\theta_{\rm sep}$$^c$}&\colhead{Origin$^a$}&\colhead{$\alpha_{B1950.0}$}& \colhead{$\delta_{B1950.0}$}\\
\colhead{(Position ID)}&\colhead{(09$^{\rm h}$51$^{\rm m}$00$^{\rm s}$)}& \colhead{(69\degr54\arcmin00\arcsec)}&\colhead{(arcsec)}&\colhead{(09$^{\rm h}$51$^{\rm m}$00$^{\rm s}$)}& \colhead{(69\degr54\arcmin00\arcsec)}
}
\startdata
\colhead{S1(42.64+58.1)} &\colhead{42.694}&\colhead{58.24}&0.31&\hii&\colhead{42.61}&\colhead{58.0} \phn \\
\colhead{S2(42.18+59.2)} &\colhead{42.210}&\colhead{59.04}&0.22& \hii&\colhead{42.19} &\colhead{58.8}\phn \\
\colhead{N1(42.39+58.2)} &\colhead{42.481}&\colhead{58.36}&0.49&\hii&\colhead{} &\colhead{}\phn \\
\colhead{N2(40.94+58.7)} &\colhead{40.938}&\colhead{58.87}&0.17&\hii&\colhead{40.97}&\colhead{58.5}\phn \\
\colhead{N3(41.20+59.6)} &\colhead{41.302}&\colhead{59.64}&0.51&SNR& &\phn \\
\enddata
\tablenotetext{a}{\citet{mcdo02}}
\tablenotetext{b}{\citet{rodr04}}
\tablenotetext{c}{Angular distance between each maser spot and the nearest continuum source.}
\end{deluxetable}

%% file: tab4.tex
\begin{deluxetable}{crrcccr}
\tabletypesize{\scriptsize} \tablecolumns{7} \tablewidth{0pc}
\tablecaption{\ho Maser in M51\label{tab:m51}} \tablehead{
\colhead{Component}&\colhead{$\alpha_{J2000.0}$}&\colhead{$\delta_{J2000.0}$}&\colhead{Peak Flux}&\colhead{Velocity}&\colhead{Feature Width\tablenotemark{a}}&
\colhead{\lwater}\\
\colhead{}&\colhead{(13$^{\rm h}$29$^{\rm m}$00$^{\rm s}$)}&
\colhead{(47\degr11\arcmin00\arcsec)}& (mJy)&\colhead{(\kmss,LSR)}&\colhead{(Channel)}&\colhead{(\lsun)}} \startdata
\colhead{Blue-shifted component}&\colhead{52.5486 $\pm$
0.0016}&\colhead{43.337 $\pm$ 0.012}&8.2&445&2&0.1\phn \\
\colhead{Red-shifted components}&\colhead{52.7089 $\pm$ 0.0001}&\colhead{42.790
$\pm$ 0.001}&41.8&539&1&0.47\phn \\ 
\colhead{}&\colhead{52.7087 $\pm$ 0.0001}&\colhead{42.789 $\pm$ 0.001}
&52.6&565&1&0.58\phn \\
\colhead{}&\colhead{52.7091 $\pm$ 0.0001}&\colhead{42.788 $\pm$ 0.002}&23.4&576&1&0.26\phn \\ 
\enddata
\tablenotetext{a}{Each channel is 5.3 \kms wide.}
\end{deluxetable}
%
%